\documentclass[a4paper,12pt]{article}
\usepackage{amsmath,amsfonts,amssymb,cite}

\begin{document}

\begin{center}
{\Large\bf Veneziano like amplitude as a test for AdS/QCD models}
\end{center}

\begin{center}
{\large S. S. Afonin}
\end{center}

\begin{center}
{\it V. A. Fock Department of Theoretical Physics, Saint-Petersburg
State University, 1 ul. Ulyanovskaya, 198504, Russia\\
E-mail: afonin@hep.phys.spbu.ru}
\end{center}

\begin{abstract}
The high energy asymptotics of QCD correlation functions is often
used as a test for bottom-up holographic models. Since QCD is not
strongly coupled in the ultraviolet domain, such a test may look
questionable. We propose that the sum over resonance poles emerging
in correlators of a bottom-up model should reproduce the structure
of a Veneziano like amplitude at zero momentum transfer assuming
equivalence of spin and radial states in the latter. This requires a
five-dimensional background that suppresses the ultraviolet part in
the effective action of a model. We give examples of emerging
low-energy holographic models.
\end{abstract}

\section{Introduction}

The ideas of the gauge/gravity correspondence from string
theory~\cite{mald,witten} have found interesting applications to
real phenomenology of the strong interactions. One example is
given by the bottom-up holographic models, called also AdS/QCD
models (the idea originated in Ref.~\cite{pol}; many references on
this activity are contained, e.g., in~\cite{forkel}). The
spectroscopy of hadrons is most successfully described within the
soft wall model introduced in Ref.~\cite{son2}. The bottom-up
approach provides a tentative semiclassical approximation to
planar QCD~\cite{hoof}.

Unfortunately, this method is not free of problems at the present
stage. One of the problems is that QCD correlators are calculated in
the deep UV domain where QCD is weakly coupled, hence, its gravity
dual should be strongly coupled in this domain. For this reason,
validity of the semi-classical approximation becomes questionable. A
straightforward possibility for improving self-consistency of the
approach consists in introducing the UV cutoff~\cite{evans,afonin}.
Such a cutoff corresponds to the scale of onset of perturbative
regime in QCD, above which the gravity dual becomes strongly
coupled. The cutoff affects both masses and residues of resonances
(since the large-$N_c$ limit is implicit in the approach, the
correlators are saturated by poles corresponding to the exchange of an
infinite number of narrow resonances~\cite{hoof}). Usually one
assumes that the spectrum of light mesons is linear in masses
squared, as predicted by hadron string models, and the available
data, in many cases, approximately agrees with this
assumption~\cite{phen} while the experimental information on
residues (related to decay constants) is very scarce except for some
ground states. Given a Regge like behavior of poles, the residues
can be fine tuned such that the high energy asymptotics of a
correlator under consideration is reproduced (see, e.g.,~\cite{we}
and references therein). As pointed out above, there is no strong
reason to expect that the QCD correlators obtained via the
holographic duality should satisfy the correct high energy
asymptotics. But then what may one use instead as a test for
calculated correlators at low energies? In the given Letter, we
propose a possible answer.

\section{Veneziano like amplitudes and resonances}

Consider, for instance, the elastic $\pi\pi$-scattering. This
process is usually described by some kind of resonance exchange. In
the large-$N_c$ limit of QCD~\cite{hoof}, the meson resonances
become narrow and one expects the appearance of an infinite number of
resonances with growing masses. At vanishing momentum transfer, the
amplitude of $\pi\pi$-scattering will be proportional to the
infinite sum over poles of $s$-channel resonances. On the other
hand, a very similar sum emerges in the resonance representation of
two-point correlation functions. This observation suggests that the
low-energy amplitudes at zero momentum transfer and the two-point
correlators could be directly related. In QCD, low energies mean the
strong coupling regime. The correlators of the strongly coupled gauge
theory in the large-$N_c$ limit may be (hopefully) obtained within
the holographic approach. This, in turn, implies that the latter
might even yield directly the scattering amplitudes at zero momentum
transfer. To motivate this suggestion further, we remind the reader
the form of the Veneziano amplitudes.

A successful ansatz for the low-energy scattering amplitudes was
proposed by Veneziano in the sixties. In particular, the most
satisfactory Veneziano like amplitude for the $\pi\pi$ scattering is
represented by the following combination of gamma
functions~\cite{collins}
\begin{equation}
\label{3}
\mathcal{A}(s,t)\sim\frac{\Gamma(1-\alpha_s)\Gamma(1-\alpha_t)}{\Gamma(1-\alpha_s-\alpha_t)};
\qquad \alpha_s=\alpha_0+\alpha's,
\end{equation}
with permutations of Mandelstam variables $s$, $t$ and $u$ that
satisfy the relation $s+t+u=4m_{\pi}^2$. We will consider the chiral
limit, $m_{\pi}=0$. The Adler self-consistency condition
(disappearance of the amplitude at vanishing momentum~\cite{collins})
leads then to the condition that $\alpha_0=\frac12$ given by the pole
$\Gamma(0)$ of denominator in~\eqref{3}. Thus, the amplitude under
consideration at $t=0$ reads
\begin{equation}
\label{4}
\mathcal{A}(s,t=0)\sim\frac{\Gamma\left(\frac12-\alpha's\right)}{\Gamma(-\alpha's)}.
\end{equation}

In the original Veneziano like amplitudes~\eqref{3}, the poles of the amplitudes
correspond to states with increasing spins $J=\alpha_s$ and the
daughter trajectories appear via extensions consisting in adding
to~\eqref{3} terms of the form
\begin{equation}
\label{12}
\Delta\mathcal{A}(s,t)\sim\frac{\Gamma(k'-\alpha_s)\Gamma(m'-\alpha_t)}{\Gamma(l'-\alpha_s-\alpha_t)},
\end{equation}
where $k',m'=1,2,\dots$ and $l'\leq k'+m'$.
At vanishing momentum transfer and $l'=1$, the generalization of the
amplitude~\eqref{4} reads
\begin{equation}
\label{13}
\mathcal{A}^{(k)}(s,t=0)\sim\frac{\Gamma\left(k+\frac12-\alpha's\right)}{\Gamma(-\alpha's)},\quad
k=0,1,2\dots.,
\end{equation}
while for $l'>1$ the generalization is
\begin{equation}
\label{13b}
\mathcal{A}^{(k,l)}(s,t=0)\sim\frac{\Gamma\left(k+\frac12-\alpha's\right)}{\Gamma(l-\alpha's)},\quad
l<k.
\end{equation}
The amplitude~\eqref{13b}, however, does not satisfy the Adler
self-consistency condition at $l>0$. We will obtain~\eqref{13b} for
the sake of completeness.

In the case of $\pi\pi$-scattering, the poles of amplitudes
correspond to states having the quantum numbers of the $\pi\pi$-system,
i.e. the spin-parities $J^P=0^{++},\,1^{--},\,2^{++},\dots$. The
position of these poles dictates a linear in masses squared spectrum
of such states. The spectrum possesses a remarkable feature: It
behaves as $m^2\sim J+n$, where $J$ is the spin and $n$ is the
radial number (the state on the $n$-th daughter trajectory at a
fixed spin). From the string point of view, this means that the
energy of oscillatory and rotational motions of a string coincide
--- such a physical picture holds for the Nambu-Goto strings which
emerged from the Veneziano model. The same spectral behaviour is
reproduced by the soft wall model~\cite{son2}. Let us assume that
the amplitudes of creation of the $J$-th and $n$-th states also
coincide, i.e., if we mark each state as $(J,n)$, the states $(J,n)$
and $(n,J)$ are completely interchangeable in the amplitude. Then
the poles in the amplitude~\eqref{4} (spin excitations at $n=0$) may
be interpreted as radial excitations with $J=0$. They must be the
same as the states saturating the two-point correlation function of
scalar currents in the large-$N_c$ limit,
\begin{equation}
\label{correl}
\langle j_S(x)j_S(0)\rangle\sim\int d^4\!x
e^{ipx}\sum_{n=0}^{\infty}\frac{Z_n}{p^2-m_n^2}.
\end{equation}
It is plausible to assume that the intermediate states in the
amplitude~\eqref{4} and in the correlator~\eqref{correl} appear with
equal probabilities (up to a normalization factor). This assumption
provides a possibility to apply the holographic approach to the
calculation of the scattering amplitudes.

Thus, our task is to reproduce the
expressions~\eqref{4},~\eqref{13}, and~\eqref{13b} within a
five-dimensional setup using the holographic method.

\section{A holographic model for the amplitude}

We first remind the reader about some basic hypotheses behind the AdS/QCD
models. According to AdS/CFT prescription~\cite{witten}, each
operator $O(x)$ of a 4D gauge theory corresponds to a field
$\Phi(x,z)$ of the 5D dual theory, with the boundary value
$\Phi(x,\varepsilon)$ being identified with the source $\Phi_0(x)$
for the operator $O(x)$. The generating functional for the
correlation functions of the 4D theory is then given by the action of the 5D
dual theory via the relation
\begin{equation}
\label{basic} W_{4D}[\Phi_0(x)]=S_{5D}[\Phi(x,\varepsilon)] \quad
\text{at} \quad \Phi(x,\varepsilon)=\Phi_0(x).
\end{equation}
Thus, if action of the 5D dual theory is known, the $n$-point
correlation functions of the 4D gauge theory can be obtained by
calculating the $n$-th functional derivative,
$\Pi_n\sim\frac{\delta^n}{\delta\Phi_0^n}S_{5D}$. It should be
emphasized that the relation~\eqref{basic} was not derived; it
represents rather a hypothesis that has passed many tests.

The AdS/QCD models are based on a bold assumption that the
holographic prescriptions may be directly applied to QCD for finding
a QCD dual theory that would describe real QCD in the strong
coupling regime. At present, it is not clear at all how to justify
this assumption because QCD is very different from the
$\mathcal{N}=4$ SYM theory in the original Maldacena's
proposal~\cite{mald}. One can try, however, to take a practical
viewpoint: Use the holographic prescriptions for building effective
models and check how good they are. The ensuing five-dimensional
models proved to represent an interesting and compact language
allowing to describe the phenomenology of QCD in the large-$N_c$
limit (short reviews are contained in~\cite{forkel}).

The original holographic duality~\cite{witten} was conjectured for
the case of anti-de Sitter (AdS) bulk space as the isometries of
AdS$_5$ are equivalent to the 4D conformal symmetry. The metric of
the AdS$_5$ space can be parametrized as
\begin{equation}
\label{2b}
ds^2=\frac{R^2}{y^2}(dx_{\mu}^2-dy^2),
\end{equation}
where $0\leq y<\infty$ and $R$ denotes the AdS radius. Since we want
to use a known holographic dictionary, it is preferable to keep the
AdS part of the 5D geometry. Our first step is to find a 5D
background that leads to the expression~\eqref{4}.

We propose the following ansatz for the 5D effective action,
\begin{equation}
\label{5} S_{5D\!,\,\text{eff}}=\int
d^4\!xdy\,\sqrt{g}\left(\frac{y}{R}\right)^3
e^{-\lambda^2y^2}(\partial_M\Phi)^2,
\end{equation}
where the AdS metric~\eqref{2b} is implied ($M=0,1,2,3,4$). Here the
dilaton like background providing the mass scale $\lambda$ is
inspired by the soft wall AdS/QCD model~\cite{son2}. The scalar
field $\Phi(x,y)$ is chosen as the simplest illustrative example. We
set $m_5=0$ for the 5D mass, the case $m_5\neq0$ will be discussed
below.

We perform now the standard steps of the bottom-up AdS/QCD models. The
equation of motion for the 4D Fourier transform $\Phi(x,y)=\int
d^4\!p\, e^{ipx}\Phi(p,y)$ is
\begin{equation}
\label{6}
\partial_y(e^{-\lambda^2y^2}\partial_y\Phi)+e^{-\lambda^2y^2}p^2\Phi=0.
\end{equation}
Evaluating the action~\eqref{5} on the solution leaves the boundary
term
\begin{equation}
\label{6b} S_{5D\!,\,\text{eff}}= \int
d^4\!x\left(\Phi\,\partial_y\!\Phi\right)_{y\rightarrow0}.
\end{equation}
Requiring that $\Phi(p,y)=\phi(p,y)\Phi_0(p)$, where $\phi(p,0)=1$,
we interpret $\Phi_0(p)$ as the Fourier transform of the source of a
scalar current corresponding to the field $\Phi(x,y)$. The solution
for $\phi(p,y)$ bounded as $y\rightarrow\infty$ reads
\begin{equation}
\phi(p,y)=\frac{1}{\sqrt{\pi}}\Gamma\left(\frac12-\frac{p^2}{4\lambda^2}\right)
U\left(-\frac{p^2}{4\lambda^2},\frac12,\lambda^2y^2\right),
\end{equation}
where $U$ denotes the Tricomi confluent hypergeometric function.

Differentiating twice with respect to the source $\Phi_0$
in~\eqref{6b} and making use of the expansion
\begin{equation}
\phi(p,y)_{y\rightarrow0}=1-2\frac{\Gamma\left(\frac12-\frac{p^2}{4\lambda^2}\right)}{\Gamma\left(-\frac{p^2}{4\lambda^2}\right)}\lambda
y + \mathcal{O}(\lambda^2y^2),
\end{equation}
we arrive at our final result,
\begin{equation}
\label{10}
\mathcal{A}(p^2)\sim
\frac{\Gamma\left(\frac12-\frac{p^2}{4\lambda^2}\right)}{\Gamma\left(-\frac{p^2}{4\lambda^2}\right)},
\end{equation}
that is identical to~\eqref{4} if we identify $p^2=s$ and
$\alpha'=(4\lambda^2)^{-1}$. The AdS radius $R$ enters the overall
factor in~\eqref{10}. In principle, given a concrete normalization
factor in the relation~\eqref{4}, it can be used for a precise
matching at $p^2=0$.

The amplitude~\eqref{10} has poles located at
\begin{equation}
p_n^2=4\lambda^2(n+1/2); \qquad n=0,1,2,\dots,
\end{equation}
which correspond to physical masses. Similarly to the soft wall
models~\cite{son2}, the same mass spectrum can be also obtained by
finding the eigenvalues of Eq.~\eqref{6} with the boundary condition
$\Phi|_{y=0}=0$. The corresponding eigenfunctions are given by the
Hermite polynomials, $\Phi_n(y)\sim H_{2n+1}(\lambda y)$.

It is easy to find a 5D background that leads to the
generalization~\eqref{13}:
\begin{equation}
\label{14} S_{5D\!,\,\text{eff}}^{(k)}=\int
d^4\!xdy\sqrt{g}\left(\frac{y}{R}\right)^{3-2k}
e^{-\lambda^2y^2}(\partial_M\Phi_{(k)})^2;\quad k=0,1,2\dots.
\end{equation}
Repeating the same steps as before we first obtain the normalized
solution of the corresponding equation of motion,
\begin{equation}
\phi_{(k)}(p,y)=\frac{\Gamma\left(k+\frac12-\frac{p^2}{4\lambda^2}\right)}{\Gamma\left(k+\frac12\right)}
U\left(-\frac{p^2}{4\lambda^2},\frac12-k,\lambda^2y^2\right),
\end{equation}
and expand this solution at $y=0$,
\begin{multline}
\phi_{(k)}(p,y)_{y\rightarrow0}=1+\sum_{i=1}^k C_i y^{2i}\prod_{j=0}^{i-1}\left(j-\frac{p^2}{4\lambda^2}\right)\\
+(\lambda
y)^{2k}\left[\frac{\Gamma\left(-k-\frac12\right)\Gamma\left(k+\frac12-\frac{p^2}{4\lambda^2}\right)}
{\Gamma\left(k+\frac12\right)\Gamma\left(-\frac{p^2}{4\lambda^2}\right)}\lambda
y + \mathcal{O}\left((\lambda y)^{3}\right)\right]. \label{b18}
\end{multline}
Here the second term yields polynomial in $y$ contributions ($C_i$
are some numerical factors). The generalization of the boundary
term~\eqref{6b} is
\begin{equation}
\label{18} S_{5D\!,\,\text{eff}}^{(k)}= \int
d^4\!x\left(\frac{\Phi_{(k)}\,\partial_y\!\Phi_{(k)}}{y^{_{2k}}}\right)_{y\rightarrow0}.
\end{equation}
Now we should differentiate~\eqref{18} twice with respect to the
source. The contributions coming from the second term in~\eqref{b18}
lead to infinities in the final answer. Since these contributions
are polynomial in $p^2$ they represent contact terms which can be
subtracted. Alternatively, one may $k+1$ times differentiate in
$p^2$ --- those terms irrelevant for physics will not survive. An
analogous situation appears when calculating the correlators of
higher-spin currents in the soft wall model. It is interesting to
note that the contact terms meet the Adler self-consistency
condition (disappearance at $p^2=0$). After subtracting we have
\begin{equation}
\label{19}
\mathcal{A}^{(k)}(p^2)\sim
\frac{\Gamma\left(-k-\frac12\right)}{\Gamma\left(k+\frac12\right)}\frac{\Gamma\left(k+\frac12-\frac{p^2}{4\lambda^2}\right)}
{\Gamma\left(-\frac{p^2}{4\lambda^2}\right)}
\end{equation}
which proves our statement. We intentionally kept the general factor
in~\eqref{19} that indicates a strong suppression of contributions
with large $k$.

To obtain the generalization of the amplitude for higher $l$, the
expression~\eqref{13b}, we must introduce the mass term
$-m^2\Phi_{(k)}^2$ into the action~\eqref{14}. It is tantamount to
the replacement $p^2\rightarrow p^2-m^2$ for the 4D momentum
squared. Setting $m^2=4\lambda^2l$, we get the expression
\begin{equation}
\label{19b} \mathcal{A}^{(k,l)}(p^2)\sim
\frac{\Gamma\left(-k-\frac12\right)}{\Gamma\left(k+\frac12\right)}\frac{\Gamma\left(k+l+\frac12-\frac{p^2}{4\lambda^2}\right)}
{\Gamma\left(l-\frac{p^2}{4\lambda^2}\right)},
\end{equation}
which has the form of~\eqref{13b}. Thus, we see that requirement
of the 5D field $\Phi$ being massless is equivalent to the Adler
self-consistency condition, $l=0$.

\section{Concluding remarks}

In comparison to the soft wall model~\cite{son2}, we have added an
extra factor to the 5D background that effectively suppresses the
weight of the high energy part of the 5D action. Speaking more
concretely, the integrand of the action of the soft wall model diverges
at $y\rightarrow0$ while we made this part finite. It is obvious
that if field $\Phi$ is a vector, the same purpose is achieved if
we replace $(\frac{y}{R})^3$ in the action~\eqref{5} by
$(\frac{y}{R})$.

A question appears as to how we should interpret the additional fields
$\Phi_{(k)}$ which result in the contributions~\eqref{19}? We
propose the following analogy. The way of inclusion of the gauge
higher spin fields, $J>1$, put forward in the soft wall
model~\cite{son2} is equivalent to considering them as coupled to
the background $\frac{ e^{-\lambda^2y^2}}{y^{2J-1}}$ after
contracting the Lorentz indices in the AdS space~\cite{afonin}. The
higher spin fields correspond to the QCD operators of higher
dimension, $\Delta=2+J$. This analogy suggests that the fields
$\Phi_{(k)}$ correspond to operators of higher dimension,
$\Delta=3+2k$. The contribution of these operators to physical
quantities is expected to be suppressed since they have higher
twist. The expression~\eqref{19} proposes an interesting estimate
for the rate of this suppression.

In summary, we have proposed a bottom-up holographic model in which
the high energy part is effectively suppressed by a 5D background.
It was conjectured that the two-point correlators in such a
low-energy model should give a structure of poles of a Veneziano
like amplitude at vanishing momentum transfer. This requirement
allows us to fix the background. The Adler self-consistency condition
turns out to require a massless field from the 5D side. We hope that
our observations may open the doors for a holographic description of
Veneziano like amplitudes.

\section*{Acknowledgments}

The work is supported by RFBR, grant 09-02-00073-a, and by the
Dynasty Foundation.

\end{document}